\documentstyle[aps,prd,eqsecnum,floats]{revtex}

\newcommand{\St}{{{\hat S}_1}}

\begin{document}
\draft

\title{ Gravitational field and equations of motion of spinning
     compact binaries to 2.5 post-Newtonian order}

\author{
Hideyuki Tagoshi
}
\address{
Department of Earth and Space Science, Osaka University,
Toyonaka, Osaka 560-0043, Japan
}

\author{
Akira Ohashi}
\address{Department of Fundamental Science, FIHS,
Kyoto University, Kyoto 606-8501, Japan}

\author{Benjamin J. Owen}
\address{
Theoretical Astrophysics 130-33, California Institute of Technology,
Pasadena, California 91125\\
and Albert Einstein Institut (Max Planck Institut f\"ur
Gravitationsphysik), Am M\"uhlenberg 1, 14476 Golm, Germany
}

\date{\today}
\maketitle

\begin{abstract}
We derive spin-orbit coupling effects on the gravitational field and
equations of motion of compact binaries in the 2.5 post-Newtonian
approximation to general relativity, one PN order beyond where spin
effects first appear.  Our method is based on that of Blanchet, Faye,
and Ponsot, who use a post-Newtonian metric valid for general
(continuous) fluids and represent pointlike compact objects with a
$\delta$-function stress-energy tensor, regularizing divergent terms
by taking the Hadamard finite part.  To obtain post-Newtonian spin
effects, we use a different $\delta$-function stress-energy tensor
introduced by Bailey and Israel.  In a future paper we will use the
2.5PN equations of motion for spinning bodies to derive the
gravitational-wave luminosity and phase evolution of binary inspirals,
which will be useful in constructing matched filters for signal
analysis.  The gravitational field derived here may help in posing
initial data for numerical evolutions of binary black hole mergers.
\end{abstract}


\section{Introduction}

Gravitational waves from coalescing compact binaries are the most
promising candidate for detection by near-future, ground-based laser
interferometers such as LIGO, VIRGO, GEO600, and TAMA \cite{Thorne}.
Detection of signals and estimation of signal parameters in noisy data
require detailed modeling of the {\it inspiral} phase of the waveform
\cite{3min} and---depending on the mass of the system---some knowledge
of the {\it merger} phase \cite{FH}.  The matched filtering techniques
used to analyze the inspiral waveform require that the model, or
template, match the real signal to less than one cycle out of the
roughly $10^3$--$10^4$ in the detectable band.  This precision
requires high-order post-Newtonian calculations of the equations of
motion and gravitational-wave luminosity.  The late inspiral and
merger stages of coalescence are not amenable to post-Newtonian
approximation methods and require numerical evolution of binary
black-hole spacetimes.  However, a post-Newtonian approximation of the
gravitational field may help in posing initial data for such numerical
evolutions~\cite{alvi}.

The standard method for modeling inspiral waveforms from binaries of
arbitrary mass ratio is the post-Newtonian expansion in powers of the
binary's orbital velocity $v/c$ and gravitational potential $GM/rc^2
\sim (v/c)^2$.  There are two nearly equivalent
approaches to calculating gravitational waves at high Post-Newtonian
order: one developed by Blanchet, Damour, and Iyer (BDI)
\cite{BandD,DI,blanchet} and one by Will and Wiseman \cite{WW} based
on previous work by Epstein, Wagoner, and Will \cite{EWW}.  The
gravitational waveforms and luminosity are expanded in time
derivatives of radiative multipoles, which are expressed as integrals
of the matter source and gravitational field.  The radiative
multipoles are combined with the equations of motion to obtain
explicit expressions in terms of the source masses, positions, and
velocities, which can be converted to gauge-invariant frequencies
observed at infinity.  Under the assumption that the bodies can be
treated as pointlike particles characterized only by their masses, the
matter source (stress-energy tensor) is given as a $\delta$-function.
In combination with a regularization scheme for infinite self-field
effects, this source greatly simplifies the field integrals compared
to a detailed fluid body calculation.

It is standard to count post-Newtonian (PN) orders in powers of
$(v/c)^2$ beyond the Newtonian result for the equation of motion and
the quadrupole formula for the gravitational waves.  For the case of
nonspinning bodies, the equations of motion have been known to 2.5PN
order for decades~\cite{OOKH,damour82,kopeikin,schafer}, and the
gravitational-wave luminosity also has been evaluated recently to
2.5PN order~\cite{blanchet}.  At the moment the 3PN calculations of
the equations of motion for nonspinning bodies are nearing completion
(see~\cite{JS,BF} and references therein).

The post-Newtonian expansion of gravitational waves from spinning
bodies has not been carried as far as for nonspinning bodies---almost
exclusively the literature contains calculations of leading-order
effects.  There are several kinds of spin effects: Spin causes
precession of the orbital plane of a binary, changes the orbital
frequency, affects the gravitational-wave luminosity, and modifies the
amplitudes of the gravitational waveforms.  Spin effects can be
further divided into spin-orbit coupling (involving a single spin) and
spin-spin interactions.\footnote{At higher post-Newtonian orders it is
also possible to have three-body, three-spin interactions; but these
have not been considered yet.}  The leading-order terms in the
equations of motion were derived in the 1950s by
Papapetrou~\cite{papa}, and in precession in the 1970s by Barker and
O'Connell~\cite{BO}.  These results have been re-derived by several
methods~\cite{damour82,DEath,IM,TH,damour,will}.  Kidder, Will, and
Wiseman~\cite{KWW} in the 1990s evaluated the leading-order (1.5PN
spin-orbit and 2PN spin-spin) terms in the gravitational-wave
luminosity (not including precession).  Kidder~\cite{Kidder} completed
that work with evaluation of the leading-order (1PN spin-orbit and
1.5PN spin-spin) terms in the waveform amplitudes.  He, along with
Apostolatos {\it et al.}~\cite{acst}, also considered the effects of
precession on the waveforms.  All of these calculations were carried
out only to leading post-Newtonian order in each spin effect, due in
large part to the unwieldiness of calculations representing spinning
bodies with a fluid stress-energy tensor of nonzero compact support.

However, it is possible to make a ``spinning particle'' approximation
to simplify the calculations.  Mino {\it et al.}  \cite{MST,TMSS} used
a $\delta$-function stress-energy tensor, a compact reformulation of
the work of Dixon \cite{dixon}, to represent a spinning test mass in
orbit around a massive spinning black hole to high post-Newtonian
order in the limit of small mass ratio.  Cho \cite{Cho} used a similar
stress-energy tensor due to Bailey and Israel \cite{BI80} to re-derive
the lowest-order results of Kidder, Will, and Wiseman~\cite{KWW}.
Recently we \cite{OTO} made the first calculation in full (arbitrary
mass ratio) post-Newtonian theory of a non-leading order effect, a 2PN
spin-orbit contribution to the waveform amplitudes, using the
stress-energy tensor of Mino {\it et al.}~\cite{MST}.  Our goal is to
eventually obtain the lowest-order unknown spin effect in the phase
evolution of the waveforms, a 2.5PN spin-orbit term.  The new term
requires evaluation of the radiative multipoles at relative 1PN order,
i.e.\ 1PN beyond the leading-order 1.5PN effect, a task which can be
done in a straightforward manner using the techniques of
Ref.~\cite{OTO}.  However there is also a contribution from the 2.5PN
spin-orbit terms in the equations of motion, which we must first
calculate.

Recently Blanchet, Faye and Ponsot (BFP)~\cite{BFP} derived the
equations of motion for nonspinning objects using a new approach,
based on the post-Newtonian fluid metric used in wave generation
calculations combined with a $\delta$-function source and the use of
the Hadamard finite part \cite{schafer,bel} to regularize the
resulting divergent integrals.  In this paper we generalize the BFP
calculation to include spin, thereby obtaining the missing 2.5PN
spin-orbit term in the equations of motion.  This term, together with
the well-known 2PN spin-spin term and the 2PN quadrupole
term~\cite{poisson,ohashi} completes the equations of motion of two
finite bodies to 2.5PN order.  The calculation is similar to our
previous work \cite{OTO}, except that we now use the stress-energy
tensor of Bailey and Israel~\cite{BI80} due to its advantages in
deriving the regularized equations of motion and precession.

We have made some effort to check that our spinning particle
approximation is actually a good model of a spinning compact object.
Of course, any $\delta$-function approximation of an extended body
yields nonsensical results near the body, but we only require a good
approximation to the field some distance outside the body and to
quantities obtained by volume integrals of the field or its source.
(The use of our field results as initial data for numerical binary
black hole simulations must be supplemented by an additional
prescription near the bodies, such as matched asymptotic
expansion~\cite{alvi}.)  Following the lead of Ref.~\cite{bel}, we
give some consistency arguments.  First, we show that the one-body
limit of our result for the gravitational field reproduces the Kerr
metric up to the post-Newtonian order considered.  Second, we verify
that the equations of motion and precession we derive from a
regularized version of the stress-energy conservation law satisfy the
harmonic gauge condition assumed at the beginning.  Third, in our
approach we reproduce the well-known 1.5PN spin-orbit terms in the
equations of motion.  Finally, we verify that the new 2.5PN terms we
derive are Lorentz invariant and reduce (in the test mass limit) to
the results of black-hole perturbation theory.

This paper is organized as follows.  In Sec.\ II, we describe our
spinning particle approximation, including the stress-energy tensor,
regularization scheme, and derivation of the regularized equations of
motion and precession.  In Sec.\ III, we use the post-Newtonian
expansion to write needed quantities in terms of a set of
post-Newtonian potentials.  Section IV is a description of the
calculation of the post-Newtonian potentials in terms of properties of
the bodies.  The (lengthy) results for the potentials are relegated to
Appendix A.  In Sec.\ V, we present the new 2.5PN terms in the
equations of motion.  In Sec.\ VI, we summarize our results and
describe problems of future interest.  In Appendix B, we check our
result for the circular-orbit frequency by comparing it to results for
spinning test particle in orbit around a Kerr black hole. In Appendix
C we provide a brief summary of the post-Newtonian check for Lorentz
invariance.

Throughout this paper, we use units such that Newton's gravitational
constant and the speed of light equal unity.  The brackets $()$ and
$[]$ on tensor indices indicate symmetrization and antisymmetrization
respectively:
\begin{equation}
\Phi_{(\alpha\beta)}=
{1\over 2}(\Phi_{\alpha\beta}+\Phi_{\beta\alpha}), \quad
\Phi_{[\alpha\beta]}=
{1\over 2}(\Phi_{\alpha\beta}-\Phi_{\beta\alpha})
\end{equation}
Greek indices run from 0 to 3, and Latin indices from 1 to 3.

\section{Spinning particle approximation}

In this section we describe how we simplify fluid-body calculations by
approximating spinning compact objects as pointlike particles.  We use
the Bailey-Israel $\delta$-function stress-energy tensor and the
Hadamard finite-part regularization scheme for the gravitational field
of a point source.  Conservation of the stress-energy tensor and
vanishing of its antisymmetric part yield the Papapetrou equations of
motion and precession in a form where the regularization is
unambiguous.

\subsection{Stress-energy tensor}

The particle-like stress-energy tensor of Bailey and
Israel~\cite{BI80} can be divided into monopole, spin, and spin
antisymmetric parts:
\begin{equation}
T^{\alpha\beta}=T^{\alpha\beta}_{(M)}+T^{\alpha\beta}_{(S)}
+T^{\alpha\beta}_{(SA)}.
\end{equation}
These parts are expressed as integrals over the trajectory $y(\tau)$
of the particle as
\begin{mathletters}
\label{e:Tmunuori}
\begin{eqnarray}
T^{\alpha\beta}_{(M)}(x)&=&
\int d\tau\, p^{\alpha}(\tau)\,u^{\beta}(\tau)
\frac{\delta^{(4)}\left(x-y(\tau)\right)}{\sqrt{-g}}
\\
T^{\alpha\beta}_{(S)}(x)&=&
- \nabla_\gamma \int d\tau
{\hat S}^{\gamma(\alpha}(\tau)u^{\beta)}(\tau)
\frac{\delta^{(4)}\left(x-y(\tau)\right)}{\sqrt{-g}},
\\
T^{\alpha\beta}_{(SA)}(x)&=&
-{1\over 2} \nabla_\gamma \int d\tau
{\hat S}^{\alpha\beta}(\tau)u^{\gamma}(\tau)
\frac{\delta^{(4)}\left(x-y(\tau)\right)}{\sqrt{-g}}.
\end{eqnarray}
\end{mathletters}
Here $u^{\mu}\equiv dy^\mu /d\tau$, $p^\mu(\tau)$ is the linear
momentum of the particle, and ${\hat S}^{\alpha\beta}(\tau)$ is an
antisymmetric tensor representing spin angular momentum.  However, the
trajectory parameter $\tau$ is generally not quite the proper time of
the body (see below).  To represent multiple bodies, we label each
particle and its associated quantities by $A$ ($y_A^\mu$, etc.) and
sum all parts of Eq.\ (\ref{e:Tmunuori}) over $A$.

When treating spinning compact bodies as point particles, we introduce
spurious degrees of freedom which must be fixed.  A finite spinning
body has a center of mass (which must be defined carefully in
post-Newtonian theory, see e.g.\ Ref.~\cite{KWW}).  However, in the
limit as the body is shrunk to a point particle, the trajectory of the
particle does not necessarily correspond to the worldline of the
finite spinning body's center of mass.  The ambiguity in precisely
what the trajectory represents (or equivalently in defining the center
of mass) is fixed by imposing a {\it spin supplementary condition.}
We choose
\begin{equation}
\label{e:SSC1}
{\hat S}^{\mu\nu}p_{\nu}=0,
\end{equation}
but note that other choices are common in the literature (see the
discussion by Kidder~\cite{Kidder}).  These other choices of spin
supplementary condition lead to equations of motion that appear
different but are physically identical, a common source of confusion.
The ambiguity in the definition of the spin also implies that $\tau$
is not precisely the proper time along the body worldline (although
$\tau$ reduces to the proper time in the Newtonian limit).  Thus
$u^\mu(\tau)$, the tangent vector to the trajectory $y^\mu(\tau)$, is
not precisely the four-velocity of the body.  We define
\begin{equation}
\hat{u}^\mu \equiv p^\mu /m,
\end{equation}
where $m$ is the Schwarzschild mass of the body.  Holding the linear
momentum constant along the worldline implies
\begin{equation}
\label{e:fixu0}
g_{\mu\nu} \hat{u}^\mu \hat{u}^\nu =-1.
\end{equation}

We also introduce a spin vector four-vector $S^\mu$.  The spin
supplementary condition (\ref{e:SSC1}) allows the natural definition
\begin{equation}
\label{e:SSC2}
{\hat S}^{\alpha\beta} = {-1\over \sqrt{-g}}
\varepsilon^{\alpha\beta\mu\nu} \hat{u}_{\mu} S_{\nu},
\end{equation}
where $\varepsilon^{\mu\nu\rho\sigma}$ is the totally antisymmetric
symbol with $\varepsilon^{0123}=1$.  Even after demanding that $S^\mu
S_\mu$ be conserved along the trajectory and that $S^i$ reduce to the
Newtonian spin in the appropriate limit, $S^0$ is still undefined.  We
fix this remaining degree of freedom by imposing the condition
\begin{equation}
\label{e:fixS0}
S^\mu p_\mu = 0.
\end{equation}
Note that the literature also contains different conventions for this
condition (see e.g.\ \cite{TH}).

\subsection{Regularization procedure}

In deriving the equations of motion, we need to evaluate the
gravitational field and associated quantities at the locations of the
bodies.  However, as usual when dealing with pointlike sources in
field theory, the quantities we need diverge if we naively integrate
the (unphysical) $\delta$-function stress-energy tensor.  Thus we must
augment our method with a regularization procedure to remove the
infinite self-field effects while preserving the correct physical
terms.  In this paper, we adopt the Hadamard finite
part~\cite{schafer,BFP,bel,BF2}.

In our calculations we encounter a class of functions $F$ which admit,
when the field point $\bbox{x}$ approaches one of the source points
($r_1= |\bbox{x} -\bbox{y}_1| \rightarrow 0$), an expansion of the
form
\begin{equation}
\label{e:had}
F(\bbox{x})=\sum_{-k_0\leq k \leq 0} r_1^kf_k(\bbox{n}_1)+O(r_1),
\end{equation}
where $k$ is a set of integers and $\bbox{n}_1 =(\bbox{x} -\bbox{y}_1)
/r_1$.  We define the effective value of the function $F$ at the
position of body 1 as the Hadamard finite part,\footnote{A more
sophisticated treatment of the Hadamard finite part is used in some
higher-order calculations~\cite{BF}. However, this version, which is
based on that of BFP, is sufficient for our purposes.}
\begin{equation}
\left(F\right)_1\equiv F(\bbox{y}_1,\bbox{y}_1)
\equiv \int {d\Omega(\bbox{n}_1)\over 4\pi} f_0(\bbox{n}_1,\bbox{y}_1).
\end{equation}
This effective value also holds for integrals of $F$ with a
three-dimensional $\delta$-function,
\begin{equation}
\int d^3\bbox{x} F(\bbox{x},\bbox{y}_1)
\delta(\bbox{x}-\bbox{y}_1)\equiv \left(F\right)_1.
\end{equation}
We introduce a symbol $()_A$ to denote the Hadamard finite part at
$x=y_A$ of the function within parentheses $()$.

When we calculate the post-Newtonian potentials, encounter integrals
of the form
\begin{equation}
\int d^3x F(x) T^{\alpha\beta}.
\end{equation}
Each part of this integral can be evaluated by means of Hadamard's
regularization as
\begin{mathletters}
\label{e:integralT}
\begin{eqnarray}
\int &&d^3x\,F(x)T^{\alpha\beta}_{(M)}(x)=
\sum_A{m_A\over u_A^0}\left(F\over \sqrt{g}\right)_A,
\label{e:integralM}
\\
\int &&d^3x\,F(x)T^{\alpha\beta}_{(S)}(x)\nonumber\\
&&
=\sum_A\left\{\left[{\hat S}_A^{\gamma(\alpha}v_A^{\beta)}
\frac{\partial_\gamma F}{\sqrt{-g}}\right]_A
-\partial_t\left({\hat S}_A^{0(\alpha}v_A^{\beta)}
\frac{F}{\sqrt{-g}}\right)_A \right.\nonumber\\
&&
\left.-\left[\left(\Gamma^\gamma_{\ \gamma\delta}
{\hat S}_A^{\delta(\alpha}v_A^{\beta)}
+{\hat S}_A^{\gamma(\alpha}\Gamma^{\beta)}_{\ \gamma\delta}
v_A^\delta\right)\frac{F}{\sqrt{-g}}\right]_A\right\},
\label{e:integralS}
\\
\int &&d^3x\,F(x)T^{\alpha\beta}_{(SA)}(x)\nonumber\\
&&
={1\over 2}\sum_A\left\{\left({\hat S}_A^{\alpha\beta}v_A^{\gamma}
\frac{\partial_\gamma F}{\sqrt{-g}}\right)_A
-\partial_t\left({\hat S}_A^{\alpha\beta}
\frac{F}{\sqrt{-g}}\right)_A \right.\nonumber\\
&&
-\left[\left(\Gamma^\alpha_{\ \mu\nu}{\hat S}_A^{\nu\beta}v_A^\mu
+\Gamma^\beta_{\ \mu\nu}{\hat S}_A^{\alpha\nu}v_A^\mu
+\Gamma^\mu_{\ \mu\nu}{\hat S}_A^{\alpha\beta}v_A^\nu\right)
\right.\nonumber\\
&&
\left.\left.\times\frac{F}{\sqrt{-g}}\right]_A\right\}.
\label{e:integralSA}
\end{eqnarray}
\end{mathletters}

\subsection{Equations of motion and precession}
\label{sec:convT}

The equations of motion and precession can be derived from the
regularized stress-energy tensor.  However, the regularization does
not commute with some derivatives and multiplications, and moreover
our version of it takes place on slices of constant $t$ rather than in
the rest frame of each body.  Therefore at each step we must make some
consistency checks.

We consider the following form of the conservation of the
stress-energy tensor,
\begin{equation}
\partial_\beta \left(\sqrt{-g} T^{\alpha\beta} \right)
+\Gamma^\alpha_{\ \beta\mu} \sqrt{-g}T^{\mu\beta}=0.
\end{equation}
By using the Bailey-Israel stress-energy tensor~(\ref{e:Tmunuori}) and
integrating over a three-volume $D_A$ containing only body $A$, we
obtain the equations of motion as
\begin{eqnarray}
\label{e:papaeq1}
{d\over dt}p_A^\alpha
+\left(\Gamma^\alpha_{\ \mu\nu}\right)_A p_A^\mu v_A^\nu
+{1\over 2}\left(R^\alpha_{\ \beta\mu\nu}\right)_A
v^{\beta}_A {\hat S}^{\mu\nu}_A=0,
\end{eqnarray}
where $v_A^\mu=u_A^\mu/u_A^0$.  (If we used Dixon's stress-energy
tensor as before~\cite{OTO}, we would have both the equations of
motion and of precession entangled in this expression, and it would be
difficult to separate them.)  By performing the same integral with the
free index covariant, we obtain
\begin{eqnarray}
\label{e:papaeq2}
{d\over dt}p_{A\alpha}
-\left(\Gamma^\mu_{\ \alpha\nu}\right)_A p_{A\mu} v_A^\nu
+{1\over 2}\left(R_{\alpha\beta\mu\nu}\right)_A
v^{\beta}_A {\hat S}^{\mu\nu}_A=0.
\end{eqnarray}
It can be shown by direct calculation that (\ref{e:papaeq1}) and
(\ref{e:papaeq2}) are equivalent up to 1PN order for non-spinning
terms and 2.5PN order for spinning terms, which is a useful
consistency check of our calculations.  Direct calculations also show
that Eqs.\ (\ref{e:papaeq1}) and (\ref{e:papaeq2}) are Lorentz
invariant up to 1PN order for non-spinning terms and 2.5PN order for
spin-orbit terms.

To derive the regularized equations of precession from the
Bailey-Israel stress-energy tensor, we demand that
\begin{equation}
\label{e:intanti}
\int_{D_A} d^3x \sqrt{-g} \,T^{[\alpha\beta]}=0.
\end{equation}
[Strictly speaking, we require that $h^{[\alpha\beta]}$ vanish, but to
the order we need this condition reduces to Eq.\ (\ref{e:intanti}).]
It is straightforward to derive the spin precession equations from
this volume integral as
\begin{equation}
\label{e:precesseq1}
{d\over dt}{\hat S}_A^{\alpha\beta}
+\left(\Gamma^\alpha_{\ \mu\nu}\right)_A
           {\hat S}^{\mu \beta}_A v^\nu_A
+\left(\Gamma^\beta_{\ \mu\nu}\right)_1
           {\hat S}^{\alpha\mu}_A v^\nu_A
+2 p_A^{[\alpha} v_A^{\beta]}=0.
\end{equation}
In the same way, we can also derive the spin precession equation with
covariant indices as
\begin{eqnarray}
\label{e:precesseq2}
{d\over dt}\left({\hat S}_{A\alpha\beta}\right)_A
+\left(\Gamma^\mu_{\ \alpha\gamma}\right)_A
           {\hat S}_{A\mu\beta} v_A^\gamma
&& +\left(\Gamma^\mu_{\ \beta\gamma}\right)_A
           {\hat S}_{A\alpha\mu} v_A^\gamma \nonumber\\
&& +2 p_{A[\alpha} v_{A\beta]}=0.
\end{eqnarray}
As we shall see in later sections, (\ref{e:precesseq1}) and
(\ref{e:precesseq2}) are equivalent, at least to the order we need.

Using Eqs.~(\ref{e:papaeq1}) and (\ref{e:precesseq1}), we can also
derive the relation
\begin{equation}
\label{e:p_and_v}
p_A^\alpha-m_A u_A^\alpha=-{1\over 2m_A}
\left(R_{\beta\nu\rho\sigma}\right)_A
S_A^{\alpha\beta}u_A^\nu S_A^{\rho\sigma}.
\end{equation}
Equation (\ref{e:p_and_v}) tells us that if we consider only
spin-orbit effects we can neglect the difference between
$\hat{u}_A^\mu$ and $u_A^\mu$.\footnote{By counting orders with the
post-Newtonian metric used in Sec.~\ref{sec3}, we see that this
difference does not matter until the 3PN spin-spin terms in the
equations of motion and the 2PN spin-spin terms in the equations of
precession.}  Therefore in the rest of this paper we set
\begin{equation}
\hat{u}_A^\mu=u_A^\mu={dy^\mu\over d\tau}
\end{equation}
and set $\tau$ to be the proper time of the body.

\section{The post-Newtonian expansion}
\label{sec3}

We introduce a parameter $\varepsilon$ of the order of a
characteristic velocity $v$.  Since we consider bound systems, we can
use the virial theorem to put $v^2=O(M/r)$ for the total mass $M$ and
(harmonic coordinate) separation $r$.  To isolate spin effects, we
also introduce a parameter $\chi$ which is also dimensionless and (at
most) of order unity for compact objects.  This $\chi$ is of order
(spin)/(mass)$^2$ for some combination of the bodies.  Terms of
$O(\chi)$ correspond to spin-orbit effects, spin-spin effects are
$O(\chi^2)$, etc.  We use the shorthand $O(m,n)$ to denote terms of
order $\varepsilon^m$ and $\chi \varepsilon^n$, and $O(m)$ to denote
terms simply of order $\varepsilon^m$ or $\chi \varepsilon^m$
(depending on the context).  In this paper we neglect terms of
$O(\chi^2)$ and higher.

We will expand numerous quantities in $\varepsilon$ and $\chi$, but
first let us determine what orders are needed.  In
Eq.~(\ref{e:papaeq1}), the Newtonian force is $O(\varepsilon^4)$; thus
the 2.5PN spin-orbit terms are $O(\chi\varepsilon^9)$.  By expanding
the entire expression in terms of the metric tensor, we find that we
need $g_{00}$ to $O(4,7)$; $g_{i0}$ to $O(5,6)$; and $g_{ij}$ to
$O(4,5)$.  By integrating different terms in the stress-energy tensor
(\ref{e:Tmunuori}) over 3-volume and substituting $S\sim \chi m^2$, we
find
\begin{eqnarray}
\label{e:orderT}
\left| T^{00} / T^{00}_{(M)} \right| &=& O(0,3), \nonumber\\
\left| T^{i0} / T^{00}_{(M)} \right| &=& O(1,2), \nonumber\\
\left| T^{ij} / T^{00}_{(M)} \right| &=& O(2,3).
\end{eqnarray}
We can insert this order counting into the metric below.

The Einstein equations can be written in harmonic coordinates as
\begin{equation}
\label{e:einstein}
\Box h^{\mu\nu} = {16\pi} |g| T^{\mu\nu} + \Lambda^{\mu\nu} (h),
\end{equation}
where
\begin{equation}
h^{\mu\nu} \equiv \sqrt{-g} g^{\mu\nu}- \eta^{\mu\nu}
\end{equation}
and the harmonic gauge condition is
\begin{equation}
\label{e:harmonic}
h^{\mu\nu}_{\ \ ,\nu} = 0.
\end{equation}
Here $\Box =\eta^{\mu\nu} \partial_\mu\partial_\nu$, $\eta^{\mu\nu} =
\rm{diag} (-1,1,1,1)$, and $\Lambda^{\mu\nu}(h)$ represents the
non-linear terms in the Einstein equations.

It is convenient to define the densities
\begin{mathletters}
\begin{eqnarray}
    \sigma &\equiv& {T^{00}+T^{ii}} \ , \\
    \sigma_i &\equiv& {T^{i0}} \ , \\
    \sigma_{ij} &\equiv& T^{ij} \ ,
\end{eqnarray}
\end{mathletters}
and the retarded potentials~\cite{BFP}
\begin{mathletters}
\begin{eqnarray}
\label{e:V}
V &=& \Box^{-1}_R\left\{-4\pi G \sigma\right\} \ ,
\\
\label{e:Vi}
V_i &=& \Box^{-1}_R\left\{-4\pi G \sigma_i\right\} \ ,
\\
\label{e:Wij}
{\hat W}_{ij} &=& \Box^{-1}_R\left\{-4 \pi G (\sigma_{ij}
- \delta_{ij} \sigma_{kk}) - \partial_i V \partial_j V\right\} \ ,
\\
\label{e:Ri}
{\hat R}_i &=& \Box^{-1}_R\left\{ - 4\pi G (V\sigma_i - V_i \sigma)
- 2 \partial_k V \partial_i V_k \right.\nonumber\\
&&\left. - {3\over 2} \partial_t V \partial_i V \right\} \ ,
\\
\label{e:X}
{\hat X} &=& \Box^{-1}_R\left\{ -4\pi G V \sigma_{ii}
+ 2 V_i \partial_t \partial_i V +V \partial_t^2 V  \right.
\nonumber \\
&&\quad\quad\left. +{3\over 2} (\partial_t V)^2
- 2 \partial_i V_j \partial_j V_i
+ \hat{W}_{ij} \partial^2_{ij} V \right\} \ ,
\end{eqnarray}
\end{mathletters}
where
\begin{equation}
\Box^{-1}_R\left\{-4\pi f( \bbox{x}, t) \right\} \equiv
\int {d^3\bbox{z}\over |\bbox{x}-\bbox{z}|}
f( \bbox{z}, t - |\bbox{x} -\bbox{z}| ).
\end{equation}
Inserting the order counting of Eq.\ (\ref{e:orderT}), we find that
the potentials have the following post-Newtonian orders:
\begin{eqnarray}
&& V=O(2,5), \quad V_i=O(3,4), \quad \hat{W}_{ij}=O(4,5), \nonumber\\
&& \hat{R}_i=O(5,6), \quad \hat{X}=O(6,7).
\end{eqnarray}

To the order we require, the solution to the Einstein
equations~(\ref{e:einstein}) is given by
\begin{mathletters}
\label{e:pnmetric}
\begin{eqnarray}
g_{00} &=& -1 + 2V - 2V^2 + 8\hat{X} + O(6,8), \\
g_{i0} &=& -4V_i - 8\hat{R}_i + O(7,8), \\
g_{ij} &=& \delta_{ij} \left( 1 + 2V + 2V^2 \right) + 4\hat{W}_{ij}
+ O(5,6),\\
\sqrt{-g} &=& 1 + 2V + 4V^2 + 2\hat{W}_{kk} + O(6,7).
\end{eqnarray}
\end{mathletters}
The harmonic coordinate condition~(\ref{e:harmonic}) reduces to the
following identities between the potentials:
\begin{mathletters}
\begin{eqnarray}
&-4\partial_t V_i-4\partial \left(\hat{W}_{ij}-
{1\over 2}\delta_{ij}\hat{W}_{kk}\right)=O(5,6),& \\
&\partial_t\left(V+{1\over 2}\hat{W}_{kk}+2V^2\right)+
\partial_i\left[V_i+2 \hat{R}_i+2V V_i\right]&\nonumber\\
&=O(6,7).&
\label{e:hident}
\end{eqnarray}
\end{mathletters}
(Actually, these are redundant and we can use only the former.)

The (regularized) post-Newtonian metric (\ref{e:pnmetric}) allows us
to relate the relativistic source parameters to the Newtonian ones.
  From here onward, indices on the right-hand sides of expressions are
raised and lowered with the Cartesian metric $\delta_{ij}$ and are
written up or down merely for convenience.  From Eq.~(\ref{e:fixu0})
we obtain
\begin{eqnarray}
u^0_A&=&
1 +{1\over2} v_A^2 +(V)_A +{3\over8} v_A^4 +{5\over2} v_A^2(V)_A
-4v_A^i \left(V_i\right)_A \nonumber\\
&&+{1\over2} \left(V^2\right)_A +O(6,7).
\end{eqnarray}
  From the definition~(\ref{e:fixS0}) of $S^0$ we obtain
\begin{equation}
S^0_A =
S^i_A \left[ v^i_A -4\left(V_i\right)_A +4v^i_A \left(V\right)_A
+O(5) \right].
\end{equation}
We introduce for convenience the shorthand $S_{A}^{ij} \equiv
\varepsilon^{ijk} S_A^k$ (where $\varepsilon^{ijk}$ is the
three-dimensional antisymmetric symbol), and by combining Eqs.\
(\ref{e:SSC1}) and (\ref{e:SSC2}) we obtain
\begin{mathletters}
\begin{eqnarray}
{\hat S}^{i0}_A&=&
S_A^{ij} \left\{ v_A^j \left[1 +{1\over2}v_A^2 +3(V)_A \right]
-4\left(V_j\right)_A \right.\nonumber\\
&&
\left. +O(5) \right\},
\\
{\hat S}^{ij}_A&=&
S^{ij}_A \left[1 +{1\over2} v^2_A -(V)_A \right] -\varepsilon_{ijk}
v^k_A (v_AS_A) \nonumber\\
&&
+m_A^2 O(\chi\varepsilon^4).
\end{eqnarray}
\end{mathletters}
Here we have also introduced the shorthand for the (Cartesian) scalar
product of two vectors $(a\,b) \equiv a^ib^i$.

\section{Calculation of the potentials}

We now describe the calculation of the potentials to the needed
orders.  The results are lengthy and are displayed in Appendix A.  Two
of the potentials ($V$ and $V_i$) are needed to relative 1PN order,
i.e.\ to O(2) beyond the leading spin and non-spin terms, while the
rest are only needed to the lowest order at which they appear.  To
obtain the explicit results in terms of the bodies' coordinates,
velocities, masses, and spins, we must substitute lower-order results
for the equations of motion and precession as well as for $V$ and
$V_i$.

The lower-order quantities we need are easily obtained.  The
lowest-order potentials are simple to evaluate as
\begin{mathletters}
\label{e:VVilowest}
\begin{eqnarray}
V&=&{m_1\over r_1}+{2 S_1^{ij}n^{i}_1 v_1^j\over r_1^2}
+1\leftrightarrow 2 +O(4,7), \\
V_i&=&{m_1\over r_1}v_1^i+{n_1^k\over 2r_1^2}S_1^{ki}
+1\leftrightarrow 2 +O(5,6),
\end{eqnarray}
\end{mathletters}
where $r_A=|\bbox{x}-\bbox{y}_A|$ and $n_A^i=(x^i-y_A^i)/r_A$.  The
lower-order terms in the equations of motion are well known.
Alternatively they can be calculated by inserting (\ref{e:VVilowest})
and the results of Sec.~III into Eq.~(\ref{e:papaeq1})
or~(\ref{e:papaeq2}) as
\begin{eqnarray}
\label{e:lowmo}
{dv_1^i\over dt} &=&
- {m_2\over r_{12}^2} \left( n_{12}^i \left\{1+ \left[ - {5m_1\over r_{12}}
- {4m_2\over r_{12}} +(v_1)^2 + 2(v_2)^2 \right.\right.\right. \nonumber\\
&&
\left.\left. - 4(v_1v_2) -{3\over 2} (n_{12}v_2)^2 \right]\right\}
- v^i_{12} \left[4(n_{12}v_1) \right. \nonumber\\
&&
\left. -3(n_{12}v_2)\right]
- \left({S_1^{ik}\over m_1}+{2S_2^{ik}\over m_2}\right)
{v_{12}^j \over r_{12}} \left(3 n_{12}^k n_{12}^j
-\delta^{kj}\right) \nonumber\\
&&
- 2\left({S_1^{kj}\over m_1}
+ {S_2^{kj}\over m_2}\right) {v_{12}^j \over r_{12}}
(3 n_{12}^k n_{12}^i-\delta^{ki}) \nonumber\\
&&
\left. +O(4,5) \right),
\label{e:1.5PNEOM}
\end{eqnarray}
with $\bbox{n}_{12}=|\bbox{y}_1-\bbox{y}_2|/r_{12}$ and
$\bbox{v}_{12}= \bbox{v}_1-\bbox{v}_2$.  The leading order terms in
the equations of precession can be calculated by the same procedure
using Eqs.\ (\ref{e:precesseq1}) and (\ref{e:precesseq2}) as
\begin{eqnarray}
\label{e:precession}
{dS_1^i\over dt}=&&{m_2\over r_{12}^2}
\left\{ S_1^i(n_{12}v_{12})
+2 n_{12}^i\left[(S_1v_2)-(S_1v_1)\right] \right. \nonumber\\
&&
\left. +(S_1n_{12})(v_1^i-2 v_2^i) + S_1\times O(4) \right\} .
\end{eqnarray}
The equations of motion and precession for body 2 are given by
exchanging labels 1 and 2 in Eqs.\ (\ref{e:1.5PNEOM}) and
(\ref{e:precession}).  Note that the effect of spin precession first
appears at 1PN order, in the sense that $dS/dt = (S\,v)/r \times
O(2)$.

The spin precession equations~(\ref{e:precession}) are different from
the usual form of the geodesic precession~\cite{TH,MTW}
\begin{eqnarray}
\label{e:precession2}
{dS_1^{\hat{i}}\over dt}=&&{m_2\over r_{12}^2}
\varepsilon_{ijk}S_1^{\hat{k}}
\varepsilon_{jpq}
(2v_2^{p}-{3\over 2}v_1^{p})n_{12}^{q}.
\end{eqnarray}
This is due to differing definitions of spin.  Equation
(\ref{e:precession2}) holds for spins defined in the local asymptotic
rest frame of each body such that $S_A^{\hat{i}} S_A^{\hat{i}}$ is
constant along the trajectory.  However, for our definition of the
spin vector this does not hold true (beyond leading order).  The spin
definitions are related by
\begin{equation}
S_1^{\hat{i}}=\left(1+{m_2\over r_{12}}\right)S_1^i
-{1\over 2}v_1^i(S_1v_1)+S_1\times O(4).
\end{equation}

The compact parts of the potentials (integrals with compact support)
are straightforward to evaluate by a retardation expansion of the
integrands of Eqs.~(\ref{e:V})--(\ref{e:X}) and substitution of the
lower-order equations of motion~(\ref{e:lowmo}) and
precession~(\ref{e:precession}).  We integrate only the symmetric part
of the stress-energy tensor.  The equations of precession
(\ref{e:precession}) ensure that $T_{(SA)}^{\alpha\beta}$ does not
contribute the metric.  It can be verified term by term that
$T_{(SA)}^{\alpha\beta}$ does not contribute to the individual
potentials either, to the order considered here.

We also encounter what are generally called quadratic terms (integrals
of products of two potentials or their derivatives) which do not have
compact support.  We use the distributional derivative
\begin{equation}
\partial^2_{ij} \left(1\over r_1\right) = {3n_1^in_1^i -\delta^{ij}
\over r_1^3} -{4\pi\over3} \delta^{ij} \delta( \bbox{x} -\bbox{y}_1),
\end{equation}
which is sufficient to our desired order~\cite{BFP}, change time
derivatives to spatial derivatives by
\begin{equation}
\partial_t \left(1\over r_1\right) = v_1^i \partial_{1i}
\left(1\over r_1\right),
\end{equation}
where $\partial_{1i} \equiv \partial/ \partial y_1^i$, use $\partial_i
(1/r_1) = -\partial_{1i} (1/r_1)$, and move the operator
$\partial_{1i}$ outside the integral sign.  We also make frequent use
of the identity
\begin{equation}
\Delta^{-1} \left(1\over r_1r_2\right) = \ln(r_1+r_2+r_{12})
\end{equation}
and its derivatives, where $\Delta$ here represents the Laplacian.

We check the potentials in two ways.  First, we verify that they
satisfy the identities (\ref{e:hident}) equivalent to the harmonic
gauge condition.  Next, we verify that the metric reduces to Kerr in
the appropriate limit.  By using the potentials in Appendix A and
setting $m_2=0$, $v_2^i=0$, $S_2^{ij}=0$, and $v_1^i=0$, we obtain the
metric
\begin{mathletters}
\begin{eqnarray}
g_{00}&=&-1+{2m_1 \over r_1}-{2m_1^2\over r_1^2}+O(6,8),
\\
g_{0i}&=&-{2\over r_1^2}n_1^k S_1^{ki}+
{2m_1\over r_1^3}n_1^k S_1^{ki}+O(6,7),
\\
g_{ij}&=&\delta_{ij}\left(1+{2m_1\over r_1}+{m_1^2\over r_1^2}
\right)+{m_1^2\over r_1^2}n_1^in_1^j+O(5,6).
\end{eqnarray}
\end{mathletters}
This is equivalent to the Kerr metric in harmonic coordinates
truncated at the indicated post-Newtonian orders.  Thus we have some
evidence that our spinning particle approximation produces physically
sensible results.

\section{Equations of motion at 2.5PN order}

Using the post-Newtonian potentials derived in the previous section,
we derive the spin-orbit interaction terms in the equations of motion
at 2.5PN order.  Direct calculation shows that it does not matter
whether we use (\ref{e:papaeq1}) or (\ref{e:papaeq2}).

\subsection{Body-centered form}

We can write the equations of motion in the Newtonian-like form
\begin{equation}
{d\over dt} {\cal P}_1^i = {\cal F}_1^i .
\end{equation}
The post-Newtonian ``linear momentum'' is given by
\begin{eqnarray}
{\cal P}_1^i &=&
m_1v_1^i \left[1 +{1\over2} v_1^2 +(V)_1 +{3\over8}v_1^4 +{5\over2}
v_1^2 (V)_1 \right. \nonumber\\
&&
\left. +{1\over2} (V^2)_1 -4v_1^j  (V_j)_1 +O(6,7) \right].
\end{eqnarray}
The ``force'' has two components, the ``frame force''
\widetext
\begin{eqnarray}
{\cal F}^i_{1(FF)} &=&
-m_1 \left( \Gamma^i_{\ \mu\nu} \right)_1 v_1^\mu v_1^\nu u^0
\nonumber\\
&=&
m_1 u_1^0 \left\{
V_{,i} \left(1+v_1^2-4V\right)
+ 4\dot{V}_i (1-2V)
+ 4V_i \left(\dot{V}+2v_1^j V_{,j}\right)
+ 8\dot{\hat{R}}_i + 4\hat{X}_{,i}
+ 8V_jV_{j,i} - 4V_{,j}\hat{W}_{ij}
- 2v_1^i\left(\dot{V} \right.
\right.\nonumber\\
&&
\left.\left.
+ v_1^jV_{,j}\right)
+ 4v_1^j \left[ 2V_{[i,j]}(1-2V) + 4\hat{R}_{[i,j]} - \dot{\hat{W}}_{ij}
\right]
+ 2v_1^jv_1^k \left(\hat{W}_{jk,i} - 2\hat{W}_{ij,k} \right)
\right\}_1
+ O(7,10)
\end{eqnarray}
and the spin-curvature coupling force
\begin{eqnarray}
{\cal F}^i_{1(SC)}&=&
-{m_1\over 2} \left( R^i_{\ \beta\mu\nu} \right)_1 v_1^\beta
\hat{S}_1^{\mu\nu} \nonumber\\
&=&
m_1 \left\{
\St^{i0}
\left( \ddot{V} + V_{,j}V_{,j} + v_1^j\dot{V}_{,j} \right)
+ \St^{ij}
\left[ \dot{V}_{,j} + 4V_{,k}V_{[j,k]} + \dot{V} V_{,j}
+ v_1^jV_{,k}V_{,k} + v_1^k \left(V_{,jk} - V_{,j}V_{,k}\right) \right]
+ \St^{j0}
\right.\nonumber\\
&&
\times \left[ V_{,ij}(1-4V) + 4\dot{V}_{(i,j)} - 3V_{,i}V_{,j}
- v_1^j\dot{V}_{,i} + 4v_1^kV_{[i,k]j} \right]
+ \St^{jk} \left[
2V_{j,ik}(1-2V) + 2\dot{\hat{W}}_{ij,k} + 4\hat{R}_{j,ik}
\right.\nonumber\\
&&
\left.\left.
- 8V_{,(i}V_{j),k} + v_1^k \left(V_{,ij} - V_{,i}V_{,j}\right)
+ 2v_1^l \left( \hat{W}_{ij,kl} +\hat{W}_{kl,ij} \right) \right]
\right\}_1 +O(7,10).
\end{eqnarray}
Note that our conventions are slightly different from those of BFP.

By inserting the explicit formulas of the post-Newtonian potentials,
given in Appendix A and by performing the Hadamard regularization, we
obtain the equations of motion in terms of body masses, spins,
positions, and velocities.  The equations of motion of body 1 can be
expressed as
\begin{eqnarray}
{dv_1^i\over dt}&=&
{m_2\over r_{12}^2}\left[A_0^i(\bbox{y}_1-\bbox{y}_2)
+A_1^i(\bbox{y}_1-\bbox{y}_2,\bbox{v}_1,\bbox{v}_2) \right.\nonumber\\
&&
+B_{1.5}^i(\bbox{y}_1-\bbox{y}_2,
\bbox{v}_1-\bbox{v}_2,\bbox{S}_1,\bbox{S}_2) \nonumber\\
&&
+B_{2}^i(\bbox{y}_1-\bbox{y}_2,\bbox{S}_1,\bbox{S}_2) \nonumber\\
&&
\left.+B_{2.5}^i(\bbox{y}_1-\bbox{y}_2,
\bbox{v}_1,\bbox{v}_2,\bbox{S}_1,\bbox{S}_2)
+O(4,6)\right]. \label{e:eomall}
\end{eqnarray}
Here $A_0^i$ and $A_1^i$ are respectively the Newtonian and 1PN forces
independent of spin, given by
\begin{eqnarray}
A_0^i&=&-n_{12}^i, \label{e:eom0pn}\\
A_1^i&=&
n_{12}^i  \left[ 5{m_1\over r_{12}} +4{m_2\over r_{12}} -v^2_1 -2v^2_2
+4(v_1v_2) \right.\nonumber\\
&&
\left.+{3\over 2} (n_{12}v_2)^2 \right]
+v^i_{12} \left[4(n_{12}v_1) -3(n_{12}v_2)\right]. \label{e:eom1pn}
\end{eqnarray}
The leading-order (1.5PN) spin-orbit interaction force is given by
\begin{eqnarray}
B_{1.5}^i&=&
\left({S_1^{ik}\over m_1}+2{S_2^{ik}\over m_2}\right)
v_{12}^j(-\delta^{kj}+3 n_{12}^k n_{12}^j)
\nonumber\\
&&+
2 \left({S_1^{kj}\over m_1}+{S_2^{kj}\over m_2}\right)
v_{12}^j(-\delta^{ki}+3 n_{12}^k n_{12}^i). \label{e:eom1.5pnSO}
\end{eqnarray}
The leading-order (2PN) spin-spin interaction force is given by
\begin{eqnarray}
B_{2}^i&=& -{3\over r_{12}^2} \left[ \bbox{n}_{12} {(S_1S_2) \over m_1m_2}
+{\bbox{S}_1 \over m_1} {(n_{12}S_2) \over m_2} +{\bbox{S}_2 \over m_2}
{(n_{12}S_1) \over m_1} \right.\nonumber\\
&&
\left. -5 \bbox{n}_{12} {(n_{12}S_1) \over m_1} {(n_{12}S_2) \over m_2}
\right]. \label{e:eom2pnSS}
\end{eqnarray}
All of these terms are well known and can be found in references such as
Damour~\cite{damour}.

\widetext
The new term, the 2.5PN (1PN beyond leading order) spin-orbit coupling
term, is given by
\begin{eqnarray}
B_{2.5}^i &=&
{n^jS_1^{ji} \over {m_1 r}} \left[ \left( {14m_1 \over r} + {9m_2 \over r}
\right) (n\,v_{12}) + {15\over2} (n\,v_{12}) (n\,v_2)^2 -3(n\,v_2)
(v_{12}v_2) \right]
+{v_{12}^jS_1^{ji} \over {m_1 r}} \left[ -{14m_1 \over r} -{9m_2 \over r}
-{15\over2} (n\,v_2)^2 \right. \nonumber\\
&&
\left. + 3(n\,v_1) (n\,v_2) - 3(v_{12}v_2) \right]
+{n^jS_2^{ji} \over {m_2 r}} \left[ \left( {35m_1 \over 2r} +{16m_2 \over r}
\right) (n\,v_{12}) -{2m_1 \over r} (n\,v_2) +15(n\,v_{12}) (n\,v_2)^2
\right. \nonumber\\
&&
\left. +6(n\,v_{12}) (v_{12}v_2) -6(n\,v_2) (v_{12}v_2) \right]
+{v_{12}^jS_2^{ji} \over {m_2 r}} \left[ -{23m_1 \over 2r} -{12m_2 \over r}
-6(n\,v_2)^2 -4(v_{12}v_2) \right]
+{n^i\over r} \left\{ {S_1^{jk} \over m_1} n^jv_{12}^k \right. \nonumber\\
&&
\times \left[ -{26m_1 \over r} -{18m_2 \over r} -15(n\,v_2)^2
-6(v_{12}v_2) \right]
+{S_2^{jk} \over m_2} n^jv_{12}^k \left[ -{49m_1 \over 2r} -{20m_2 \over
r} -15(n\,v_2)^2
-6(v_{12}v_2) \right]
\nonumber\\
&&
\left.
  -12\varepsilon^{jkl} n^jv_1^kv_2^l
{(v_1S_1) \over m_1} \right\}
+ {v_{12}^i \over r}\left\{ {S_1^{jk} \over m_1} \left[ -6(n\,v_1) n^jv_{12}^k
\right]
+{S_2^{jk} \over m_2} \left[ -6(n\,v_1) n^jv_{12}^k \right] \right\}
+ {v_1^i\over r} \left\{ {S_1^{jk} \over m_1} \left[ 3(n\,v_{12}) n^jv_1^k
\right.\right.\nonumber\\
&&
\left.\left.
-3v_1^jv_2^k \right]
+{S_2^{jk} \over m_2} \left[ 6(n\,v_{12}) n^jv_2^k
-4v_1^jv_2^k \right] \right\}
+ {\varepsilon^{ijk}\over r} \left\{ -n^jv_{12}^k {4m_1 \over r} {(n\,S_2)
\over m_2}
-3n^jv_{1}^k(n\,v_{12}) {(v_1S_1) \over m_1}
\right.\nonumber\\
&&
\left.
-6n^jv_{2}^k(n\,v_{12}) {(v_1S_1) \over m_1}
+7v_1^jv_2^k {(v_1S_1) \over m_1}\right\}. \label{e:eom2.5pnSO}
\end{eqnarray}

\subsection{Center-of-mass form}

Here we express the equations of motion in terms of the bodies'
relative coordinate, in the center-of-mass frame.  We define the mass
parameters $M=m_1+m_2$, $\eta=m_1 m_2/M^2$, and $\Delta=(m_1-m_2)/M$;
and the dimensionless, symmetrized spin parameters~\cite{WW}
\begin{mathletters}
\begin{eqnarray}
\bbox{\chi}_s&=&{1\over 2}
\left({\bbox{S}_1\over m_1^2}
+{\bbox{S}_2\over m_2^2}\right),\\
\bbox{\chi}_a&=&{1\over 2}
\left({\bbox{S}_1\over m_1^2}
-{\bbox{S}_2\over m_2^2}\right).
\end{eqnarray}
\end{mathletters}
It is also convenient to introduce the shorthand
\begin{mathletters}
\begin{eqnarray}
\chi_s^{ij}&=&{1\over 2}\left({S_1^{ij}\over m_1^2}
+{S_2^{ij}\over m_2^2}\right), \\
\chi_a^{ij}&=&{1\over 2}\left({S_1^{ij}\over m_1^2}
-{S_2^{ij}\over m_2^2}\right).
\end{eqnarray}
\end{mathletters}
The relation between the body coordinates $\bbox{y}_1$, $\bbox{y}_2$,
and the relative coordinate is
\begin{eqnarray}
\label{e:com}
\bbox{y}_1&=&\bbox{x}\left[{m_2\over M}+{1\over 2}\eta
\Delta\left(v^2-{M\over r}\right)\right]
-M\eta\bbox{v}\times(\bbox{\chi}_a+\Delta\bbox{\chi}_s),
\nonumber\\
\bbox{y}_2&=&\bbox{x}\left[-{m_1\over M}+{1\over 2}\eta
\Delta\left(v^2-{M\over r}\right)\right]
-M\eta\bbox{v}\times(\bbox{\chi}_a+\Delta\bbox{\chi}_s).
\end{eqnarray}
The last terms in the above equation are 1.5PN order and contribute to
the 2.5PN relative acceleration through corrections to $A_1^i$.  From
now on we drop the subscript ${\scriptstyle 12}$ on $r$, $\bbox{n}$,
and $\bbox{v}$.

We define the relative acceleration as
\begin{equation}
\bbox{a}\equiv {d\bbox{v}_1\over dt}-
{d\bbox{v}_2\over dt},
\end{equation}
and we express the equations of motion as
\begin{equation}
\label{e:eomrelative}
\bbox{a}=\bbox{a}_N+\bbox{a}_{PN}+\bbox{a}_{SO}
+\bbox{a}_{SS}+\bbox{a}_{PNSO}+O(6,8).
\end{equation}
The Newtonian acceleration $\bbox{a}_N$ is given by
\begin{equation}
\label{e:eomrelaN}
\bbox{a}_N=-{M\over r^2}\bbox{n}.
\end{equation}
The 1PN acceleration $\bbox{a}_{PN}$ is given by
\begin{eqnarray}
\label{e:eomrela1pn}
\bbox{a}_{PN}&=&-{M\over r^2}
\left\{ \bbox{n} \left[ (1+3\eta) v^2 -2(2+\eta) {M\over r}
-{3\over 2}\eta (n\,v) \right] \right. \nonumber\\
&&
\left. -2(2-\eta) (n\,v)\bbox{v} \right\}.
\end{eqnarray}
The 1.5PN spin-orbit acceleration $\bbox{a}_{SO}$ is given by
\begin{eqnarray}
\label{e:eomrela15so}
\bbox{a}_{SO}&=&{M^2\over r^3}
\left\{6\bbox{n}(\bbox{n}\times \bbox{v})
\cdot (\bbox{\chi}_s+\Delta\bbox{\chi}_a)\right.
\nonumber\\
&&
-2\bbox{v}\times[(2-\eta)\bbox{\chi}_s+2\Delta\bbox{\chi}_a]
\nonumber\\
&&\left.
+6(n\,v)\bbox{n}\times[(1-\eta)\bbox{\chi}_s+\Delta \bbox{\chi}_a]
\right\}.
\end{eqnarray}
The 2PN spin-spin acceleration $\bbox{a}_{SS}$ is given by
\begin{eqnarray}
\bbox{a}_{SS}&=&
-{M^3\over r^4} 3\eta \left\{
\bbox{n}
\left[ |\bbox{\chi}_s|^2-|\bbox{\chi}_a|^2-5(n\,\chi_s)^2
+5(n\,\chi_a)^2 \right] \right. \nonumber\\
&&
\left. +2[\bbox{\chi}_s(n_{12}\chi_s) -\bbox{\chi}_a(n\,\chi_a)]
\right\}.
\end{eqnarray}

\widetext
The new 2.5PN spin-orbit acceleration $\bbox{a}_{PNSO}$ (i.e.\ the
post-Newtonian correction to $\bbox{a}_{SO}$) is given by
\begin{eqnarray}
\label{e:eomrela25so}
a_{PNSO}^i &=& {M^2 \over 2r^3} \left( \chi_s^{ji} \left\{ n^j (n\,v)
\left[ \left(32 -15\eta +8\eta^2 \right) {M\over r} +15\eta (1+2\eta)
(n\,v)^2 -6\eta (3+2\eta) v^2 \right]
+v^j \left[ 3(-8+\eta) {M\over r} \right.\right.\right. \nonumber\\
&&
\left.\left. -3\eta (5+2\eta) (n\,v)^2 +14\eta v^2 \right] \right\}
+\Delta\chi_a^{ji} \left\{ n^j(n\,v) \left[ (32+13\eta) {M\over r}
+15\eta (n\,v)^2 -18\eta v^2 \right]
+v^j \left[ -3(8+3\eta) {M\over r} \right.\right. \nonumber\\
&&
\left.\left. -15\eta (n\,v)^2 +14\eta v^2 \right] \right\}
+n^in^jv^k \left\{ \chi_s^{jk} \left[ \left(-40 -13\eta -16\eta^2
\right) {M\over r} -30\eta (n\,v)^2 +24\eta v^2 \right]
+\Delta\chi_a^{jk} \right. \nonumber\\
&&
\left.\times \left[ (-40-21\eta) {M\over r} -30\eta (n\,v)^2
+24\eta v^2 \right] \right\}
-6v^in^jv^k (n\,v) \left[ \left(2 -3\eta -2\eta^2 \right) \chi_s^{jk}
+(2-\eta) \Delta\chi_a^{jk} \right]
+(\bbox{n} \times \bbox{v})^i \nonumber\\
&&
\left. \left\{ -8\eta [(n\,\chi_s) +\Delta (n\,\chi_a)] {M\over r}
-6\eta (n\,v) [(1+2\eta) (v\,\chi_s) -\Delta (v\,\chi_a)] \right\}
\right).
\end{eqnarray}

\subsection{Quasicircular orbits}

When the spins are aligned or anti-aligned with the orbital angular
momentum, the spin vectors and the orbital angular momentum vector do
not precess.  In this case, a quasicircular orbit exists.  We define
$\chi_a=\pm |\bbox{\chi}_a|$ and $\chi_s=\pm |\bbox{\chi}_s|$.  The
signs of $\chi_a$ and $\chi_s$ are positive when the vectors are
aligned with the angular momentum axis $(\bbox{n}\times\bbox{v})$ and
negative when antialigned.  For quasicircular orbits, the equations of
motion take the form
\begin{equation}
\bbox{a}=-\omega^2 \bbox{x},
\end{equation}
where $\omega$ is the orbital angular frequency.  Using the identity
$v=\omega r$, we can eliminate $v$ from the equations of motion and
express $\omega$ as an expansion in $M/r$,
\begin{eqnarray}
\label{e:omegaH}
\omega^2&=&
{M\over r^3} \left\{ 1 -(3-\eta) {M\over r} -2\left[ (1+\eta) \chi_s
+\Delta\chi_a \right] \left( {M\over r} \right)^{3/2} \right.\nonumber\\
&&\left.
+3\eta \left(M \over r\right)^2 \left[ (\chi_s)^2 -(\chi_a)^2 \right]
+\left[ (9-3\eta+\eta^2) \chi_s \right.\right. \nonumber\\
&&
\left.\left. +(9-6\eta) \Delta\chi_a \right] \left(M \over r\right)^{5/2}
+O(4,7) \right\}.
\end{eqnarray}
In Appendix B, we show that Eq.~(\ref{e:omegaH}) agrees in the limit
$\eta\to0$ with the motion of a test particle around a Kerr black hole
to 2.5PN order.

The new terms in Eq.~(\ref{e:omegaH}) are proportional to $M/r^3(M/r)^{5/2}$:
The $O(\eta^2)$ term and parts of the $O(\eta)$ terms have not been derived
before.  Their magnitude indicates that 2.5PN spin-orbit contributions change
the circular orbit frequency by an amount of less than about two percent near
the innermost stable orbit for a system of maximally spinning black holes.
However, the importance of this effect on the phase evolution of the orbit, and
thus on the effectiveness of matched filtering for the gravitational-wave
signal, cannot be evaluated until the gravitational-wave luminosity is known.

\section{Summary}

We have calculated the 2.5PN spin-orbit effects on the gravitational
field and equations of motion of compact binaries by approximating the
compact bodies as spinning point particles.  Although the formal basis
of this approximation remains uncertain, we have demonstrated that it
is reasonable in that it reproduces some physical behavior of the Kerr
spacetime to this post-Newtonian order, as well as lower-order,
previously known terms in the equations of motion.

Using the techniques of this paper, it is a straightforward though
lengthy task to calculate the equations of precession to 2PN order and
the gravitational-wave luminosity (needed to construct templates) to
2.5PN order.  We plan to do this in the future for circular orbits.
The 2.5PN gravitational field given by combining our potentials with
the spinless potentials given by BFP could serve toward providing
initial data for the numerical evolution of spinning binary black-hole
mergers, if augmented by some scheme for approximating the field near
and inside the black holes~\cite{alvi}.

\section*{Acknowledgments}

We thank Guillaume Faye, Misao Sasaki, Takahiro Tanaka, Kip Thorne,
and especially Luc Blanchet for helpful discussions.  HT also thanks
Kip Thorne and Luc Blanchet for warm hospitality at Caltech and DARC
Observatoire de Paris-Meudon, and thanks the Albert Einstein Institut
for hospitality.  HT and AO were supported by the Japanese Society for
the Promotion of Science.  HT and BJO were supported by NSF Grant
PHY-9424337 and NASA Grant NAG5-6840.  HT was also supported by CNRS
of France and by Monbusho Grant-in-Aid 11740150, and BJO was also
supported by the NSF Graduate Program.  We performed numerous
algebraic manipulations with the Mathematica and MathTensor software
packages.

\appendix

\section{Explicit post-Newtonian potentials}

In this appendix, we present explicit expressions (in terms only of
the masses, spins, positions, and velocities) for the post-Newtonian
potentials necessary to obtain the equations of motion and the metric
away from the bodies.  We denote spin and monopole parts of
quantities with subscript $(S)$ and $(M)$, respectively: e.g.,
$V_{(S)}$ and $V_{(M)}$.  All expressions are fully reduced using the
equations of motion (\ref{e:1.5PNEOM}) and precession
(\ref{e:precession}).  The monopole terms in the potentials are given
to higher order by BFP, but for convenience we reproduce the pieces
needed in our calculation.

\widetext
The relevant potentials are
\begin{mathletters}
\begin{eqnarray}
V_{(M)}&=&{m_1\over r_1}+m_1\left[{2 v_1^2 \over r_1}-
{(n_1 v_1)^2 \over 2 r_1}-{m_2 r_1\over 4 r_{12}^3}
-{5m_2\over 4 r_1 r_{12}}
+{m_2 r_2^2 \over 4 r_1 r_{12}^3}\right]
+1\leftrightarrow 2+O(5),
\end{eqnarray}
\begin{eqnarray}
V_{(S)}&=&
S_1^{ij} \left( n_1^i v_1^j \left[ {2\over r_1^2} + {2v_1^2 \over r_1^2}
- {3\over r_1^2} (n_1v_1)^2 - {2m_2\over r_1^2 r_{12}}
+{m_2\over r_1 r_{12}^2} (n_1n_{12}) +{5\over2} {m_2\over r_{12}^3} \right]
-n_1^i v_2^j \left( {4m_2\over r_1^2 r_{12}} +{5\over 2}{m_2\over r_{12}^3}
\right) +n_1^i n_{12}^j \left[ {2m_2\over r_1 r_{12}^2} \right.\right.
\nonumber\\
&&
\left.\times (n_1v_1) -{9m_2\over 2r_{12}^3} (n_{12}v_{12}) \right]
+n_{12}^iv_1^j \left\{ {7m_2\over r_1r_{12}^2}
-{3m_2\over r_{12}^3} [(n_1n_{12}) -(n_2n_{12})] \right\}
-n_{12}^iv_2^j \left\{ {4m_2\over r_1r_{12}^2} -{3m_2\over r_{12}^3}
[(n_1n_{12}) \right. \nonumber\\
&&
\left.\left. -(n_2n_{12})] \right\}
-2n_{12}^iv_{12}^j{m_2\over r_2r_{12}^2}
-2n_2^i v_{12}^j {m_2\over r_{12}^3}
+3n_2^i n_{12}^j (n_{12}v_{12}) {m_2\over r_{12}^3} \right)
+1\leftrightarrow 2 + O(8),
\end{eqnarray}
\begin{eqnarray}
V_{i(M)}=&& {m_1 v_1^i \over r_1}
\left[1+v_1^2-{1\over 2}(n_1 v_1)^2\right]
+m_1 m_2\left[{(n_1 v_1)\over r_{12}^2}n_{12}^i
+{3\over 2}{r_1\over r_{12}^3}n_{12}^i
(n_{12}v_{12})
+{1\over 2}{r_1\over r_{12}^3}v_2^i \right]
\nonumber
\\
&&
+{m_1 m_2 \over 4}v_1^i
\bigg(-{3 r_1\over r_{12}^3}-{5 \over r_1r_{12}}
+{r_2^2\over r_1r_{12}^3}\bigg)
+1\leftrightarrow 2+O(6),
\end{eqnarray}
\begin{eqnarray}
V_{i(S)}&=&
S_1^{ji} \left\{ n_1^j \left[ {1\over 2r_1^2} +{v_1^2\over 2r_1^2}
-{3 m_2\over 2r_1^2 r_{12}} -{3\over 4r_1^2} (n_1v_1)^2 +{m_2\over 4r_1
r_{12}^2} (n_1 n_{12}) \right] +n_{12}^j {3m_2\over 4r_1 r_{12}^2} \right\}
+ v_1^i S_1^{jk} {n_1^jv_1^k \over 2r_1^2} +\varepsilon^{ijk} n_1^jv_1^k
{(v_1 S_1) \over 2r_1^2} \nonumber\\
&&
+1\leftrightarrow 2 +O(7),
\end{eqnarray}
\begin{eqnarray}
{\hat W}_{ij(M)} &=&
\delta^{ij} \left(  -\frac{ m_1 v_1^2}{r_1}-\frac{m_1^2}{4 r_1^2}+
\frac{m_1 m_2}{r_{12} s} \right)
+\frac{m_1 v_1^i v_1^j}{r_1} +\frac{m_1^2 n_1^i n_1^j}{4 r_1^2}
+m_1 m_2 \left[ \frac{1}{s^2} \left(n_1^{(i} n_2^{j)}
+ 2n_1^{(i} n_{12}^{j)} \right)-
n_{12}^i n_{12}^j \left(\frac{1}{s^2}
\right.\right.\nonumber\\
&&
\left.\left.
+\frac{1}{r_{12} s} \right) \right]
+1\leftrightarrow 2 + O(5),
\end{eqnarray}
\begin{equation}
{\hat W}_{ij(S)}=
\left( S_1^{k(i}v_1^{j)} -\delta_{ij} S_1^{kl}v_1^{l} \right)
{n_1^k\over r_1^2} +1\leftrightarrow 2 +O(6),
\end{equation}
\begin{eqnarray}
{\hat R}_{i(M)}=&&
m_1 m_2 n_{12}^i \left[ -\frac{(n_{12}v_1)}{2s} \left(\frac{1}{s}+
\frac{1}{r_{12}} \right)
-\frac{2 (n_2v_1)}{s^2}+\frac{3(n_2v_2)}{2s^2} \right]
+ n_1^i \left\{ \frac{m_1^2 (n_1v_1)}{8 r_1^2}
+\frac{m_1 m_2}{s^2} \left[ 2 (n_{12}v_1) - \frac{3}{2}(n_{12}v_2)
\right.\right.\nonumber\\
&&
\left.\left.
+2 (n_2v_1) - \frac{3}{2}(n_2v_2) \right] \right\}
+v_1^i \left[ -\frac{m_1^2}{8r_1^2}
+ m_1 m_2 \left( \frac{1}{r_1 r_{12}}
+ \frac{1}{2 r_{12}s} \right) \right]
-v_2^i\frac{m_1 m_2}{r_1 r_{12}}.
+1\leftrightarrow 2+O(6),
\end{eqnarray}
\begin{eqnarray}
{\hat R}_{i(S)}&=&
S_1^{ji} \left[ n_1^j \left( -{m_1\over 4r_1^3} +{m_2\over 2r_1^2r_{12}}
+{m_2\over r_{12}s} \right) +n_{12}^j \left( -{m_2\over 2r_1r_{12}^2}
+{m_2\over 2r_{12}^2r_2} +{m_2\over r_1s^2} \right) +n_2^j \left(
{m_2\over r_1s^2} +{m_2\over r_{12}s^2} \right) \right] +n_1^iS_1^{jk}
\left[ n_1^j \right. \nonumber\\
&&
\left.\times \left(n_{12}^k +n_2^k\right) \left( {m_2\over r_1s^2}
+{2m_2\over s^3} \right) -2n_{12}^jn_2^k {m_2\over s^3} \right]
+n_{12}^iS_1^{jk} \left[ \left(n_1^j +n_2^j\right) n_{12}^k \left(
-{m_2\over r_{12}s^2} -{2m_2\over s^3} \right) -2n_1^jn_2^k {m_2\over
s^3} \right] \nonumber\\
&&
+1\leftrightarrow 2 +O(7),
\end{eqnarray}
\begin{eqnarray}
{\hat X}_{(S)}&=&
S_1^{ij} \left\{ n_1^iv_1^j \left( -{m_1\over 2r_1^3} +{m_2\over
r_1^2r_{12}} +{m_2\over r_{12}s^2} +{m_2\over r_2s^2} \right)
+n_{12}^iv_1^j \left( -{m_2\over r_1r_{12}^2} +{m_2\over r_{12}^2r_2}
-{m_2\over r_2s^2} \right) +n_2^iv_1^j {m_2\over r_{12}s^2}
+n_1^iv_2^j \left( {m_2\over r_{12}s^2} \right.\right. \nonumber\\
&&
\left. -{m_2\over r_2s^2} \right)
+n_{12}^iv_2^j \left( {2m_2\over r_1s^2} +{m_2\over r_2s^2} \right)
+n_2^iv_2^j \left( {2m_2\over r_1s^2} +{m_2\over r_{12}s^2} \right)
+n_1^in_{12}^j \left[ (n_{12}v_1) \left( -{m_2\over r_{12}s^2}
-{2m_2\over s^3} \right) -2(n_2v_{12}) {m_2\over s^3} \right.
\nonumber\\
&&
\left. +(n_1v_2) \left(
{2m_2\over r_1s^2} +{4m_2\over s^3} \right) +(n_{12}v_2) \left(
-{m_2\over r_1s^2} -{2m_2\over s^3} \right) \right] +n_1^in_2^j \left[
(n_1v_2) \left( {2m_2\over r_1s^2} +{4m_2\over s^3} \right)
-2(n_{12}v_1) {m_2\over s^3} \right. \nonumber\\
&&
\left. +(n_2v_{12}) \left( -{m_2\over r_2s^2}
-{2m_2\over s^3} \right) -2(n_{12}v_2) {m_2\over s^3} \right]
+n_{12}^in_2^j \left[ (n_{12}v_1) \left( {m_2\over r_{12}s^2}
+{2m_2\over s^3} \right) +(n_2v_{12}) \left( {m_2\over r_2s^2}
+{2m_2\over s^3} \right) \right. \nonumber\\
&&
\left.\left. +(n_{12}v_2) \left( {m_2\over r_{12}s^2}
+{2m_2\over s^3} \right) -4(n_1v_2) {m_2\over s^3} \right] \right\}
+1\leftrightarrow 2+O(8),
\end{eqnarray}
\end{mathletters}
where $s=r_1+r_2+r_{12}$. These potentials, inserted into the
post-Newtonian metric~(\ref{e:pnmetric}), give the spin-orbit
interaction terms in the gravitational field to 2.5PN order away from
the bodies.

Using the Hadamard finite part, we find the values of the spin parts
of the potentials at body 1 to be
\begin{mathletters}
\begin{eqnarray}
\left( V \right)_{1(S)} &=&
{1\over r_{12}^2} \left( -S_1^{pq} {m_2 \over r_{12}} n_{12}^p v_{12}^q
+S_2^{pq} n_{12}^p \left\{ \left[2
-{21m_1 \over 2r_{12}}
\right.\right.\right.
\nonumber\\
&&
\left.\left.\left.
+2v_2^2 -3(n_{12}v_2)^2 \right]v_2^q
+{m_1 \over 2r_{12}} v_1^q \right\} \right)
\nonumber\\
&&
+O(8),
\\
\left( V_i \right)_{1(S)} &=&
{1\over r_{12}^2} \left\{ S_2^{pi} n_{12}^p \left[ {1\over2} -{5m_1
\over 2r_{12}} +{1\over2} (v_2)^2 \right.\right. \nonumber\\
&&
\left. -{3\over4} (n_{12}v_2)^2 \right] +{1\over2} v_2^i S_2^{pq}
n_{12}^p v_2^q \nonumber\\
&&
\left. +{1\over2} \epsilon^{ipq} n_{12}^p v_2^q (v_2S_2) \right\}
+O(7), \\
\left( \hat{W}_{ij} \right)_{1(S)} &=&
{1\over r_{12}^2} \left( S_2^{p(i} v_2^{j)} n_{12}^p -\delta^{ij}
S_2^{pq} n_{12}^p v_2^q \right)
\nonumber\\
&&
+O(6), \\
\left( \hat{R}_i \right)_{1(S)} &=&
{1\over r_{12}^2} n_{12}^p \left[ S_1^{pi} {m_2 \over 4r_{12}}
+S_2^{pi} \left( {m_1 \over r_{12}} -{m_2 \over 4r_{12}} \right)
\right] \nonumber\\
&&
+O(7),\\
\left( \hat{X} \right)_{1(S)} &=&
{1\over r_{12}^2} \left\{ S_1^{pq} n_{12}^p \left( v_1^q -{1\over2}
v_2^q \right) {m_2 \over r_{12}} +S_2^{pq} n_{12}^p \right.\nonumber\\
&&
\left.\times \left[ {m_1 \over 4r_{12}} v_1^q + \left( {7m_1 \over
4r_{12}} -{m_2 \over 2r_{12}} \right) v_2^q \right] \right\}
\nonumber\\&&+O(8).
\end{eqnarray}
\end{mathletters}

\section{Comparison with Kerr orbital frequency}

We perform one check of our equations of motion by comparing the
orbital frequency (\ref{e:omegaH}) in the test-mass limit to the
orbital frequency of a spinning test particle in a circular orbit
around a Kerr black hole derived by Tanaka {\it et al.} \cite{TMSS}

The transformation from Boyer-Lindquist coordinates $(t_{BL}, r_{BL},$
$\theta_{BL}, \phi_{BL})$ to harmonic coordinates $(t_H, x_H,$
$y_H, z_H)$ was found by Cook and Scheel \cite{CS} to be
\begin{mathletters}
\label{e:BLtoH}
\begin{eqnarray}
t_H&=&t_{BL}+{{r_+^2+a^2}\over {r_+-r_-}}
\ln\left|{{r_{BL}-r_+}\over {r_{BL}-r_-}}\right|,
\\
x_H+iy_H&=&(r_{BL}-M+ia)e^{i\phi}\sin\theta_{BL},
\\
z_H&=&(r_{BL}-M)\cos\theta_{BL},
\end{eqnarray}
\end{mathletters}
where
\begin{equation}
\phi=\phi_{BL}+{a\over {r_+-r_-}}
\ln\left|{{r_{BL}-r_+}\over {r_{BL}-r_-}}\right|,
\end{equation}
$M$ and $a$ are the Kerr mass and spin parameters, and
$r_{\pm}=M\pm\sqrt{M^2-a^2}$.  The angular frequency in
Boyer-Lindquist coordinates is the same as in harmonic coordinates,
$\omega= d\phi_{BL}/ dt_{BL}= d\phi_H/ dt_H$, where
$\phi_H=\tan^{-1}(y_H/x_H)$.  The transformation of
the radial coordinate can be written as
\begin{equation}
\label{e:trans}
r_{H}=r_{BL}-M+O(a^2),
\end{equation}
which is sufficiently accurate for spin-orbit effects.

The orbital angular frequency $\omega$ of a test particle in a
circular orbit in the Kerr spacetime with spin perpendicular to the
plane of the orbit is given by Eq.\ (4.26) of Ref.\ \cite{TMSS}.  That
equation can be written as
\begin{equation}
\omega
={M_{BH}^{1/2}\over r_{BL}^{3/2}}
\left[1 -\left(q +{3\over2} {m_p\over M_{BH}} s_\perp \right)
\left(M_{BH} \over r_{BL}\right)^{3/2} +O(6) \right].
\end{equation}
Here $q=a/M$, $s_\perp$ is the magnitude of the spin of the particle
divided by the square of its mass, $m_p$ is the particle's mass, and
$M_{BH}$ is the (much greater) black hole's mass.  Note that the
effect of the particle's spin $s_\perp$ is already $O(m_p/M_{BH})$
smaller than that of the black hole.\footnote{The coordinate
transformation (\ref{e:trans}) is modified by the presence of the
particle.  However the modification appears at $O(m_p/M_{BH})$ in the
spinless terms and at $O(m_p/M_{BH})^2$ in the spin terms; thus it
does not affect the comparison here.}  The parameter $s_\perp$ is
explicitly written in terms of the coordinate components of the spin
vector as
\begin{equation}
s_\perp={S_p^\theta\over m_p^2}\sqrt{r_{BL}^2+a^2\cos\theta_{BL}},
\end{equation}
where $S_p^\theta$ is the $\theta_{BL}$-component of the spin vector
of particle.  Since we consider the equatorial plane, $\theta=\pi/2$.
Using the coordinate transformation (\ref{e:BLtoH}), we find that
\begin{equation}
s_\perp=\left(1+{M_{BH}\over r_H}\right){S_2^z\over m_2^2},
\end{equation}
where $S_2^z$ is the $z$-component of the spin vector of the particle
in harmonic coordinate and $m_2=m_p$ is the particle's mass.

Using above formulas, we write the angular frequency of a test particle
in harmonic coordinates as
\begin{eqnarray}
\omega^2&=&
\left(M_{BH}\over r_{H}^{3}\right) \left[1 -3\left(M_{BH}\over
r_H\right) -\left(2q +3{m_2\over M_{BH}} s_{2} \right) \right.\nonumber\\
&&
\times \left(M_{BH}\over r_H\right)^{3/2} +6\left(M_{BH}\over
r_H\right)^2 +\left(9q +{21\over2} {m_2\over M_{BH}} s_{2}\right)
\nonumber\\
&&
\left. \times\left(M_{BH}\over r_H\right)^{5/2} +O(4,6)\right],
\end{eqnarray}
where $s_2\equiv S_2^z/m_2^2$.
By taking our post-Newtonian result (\ref{e:omegaH}) for $\omega^2$,
setting body 1 to be the Kerr black hole and body 2 to be the particle,
taking the test-mass limit
\begin{eqnarray*}
\eta&\approx& {m_2\over M_{BH}}, \\
\Delta&\approx&1-2{m_2\over M_{BH}}, \\
\chi_s&=&{1\over 2}\left(q+s_{2}\right), \\
\chi_a&=&{1\over 2}\left(q-s_{2}\right),
\end{eqnarray*}
we obtain the same result.  Thus our post-Newtonian calculation agrees
(in the test-mass limit) with the previous results of black-hole
perturbation theory.

\section{Lorentz invariance of the equations of motion}

In this appendix, we briefly describe the Lorentz invariance of the
equations of motion (\ref{e:eomall}).  It is well known that the 2PN
equations of motion, including spin-orbit and spin-spin terms, are
Lorentz invariant~\cite{damour}.  Therefore we concentrate on the new
2.5PN spin-orbit terms.

Consider the case where the coordinates $(t',x^{i'})$ are moving
relative to $(t,x^i)$ with constant velocity $U^i$.  The Lorentz
transform of a vector $A$ is defined as
\begin{equation}
\label{ulorentz}
A^\mu = \Lambda^\mu{}_{\nu'} A^{\nu'} ,
\end{equation}
where
\begin{mathletters}
\begin{eqnarray}
\Lambda^0{}_0 &=& \gamma , \\
\Lambda^0{}_{i'} &=& \Lambda^i{}_{0'} = \gamma U^i , \\
\Lambda^i{}_{j'} &=& \delta^{ij}+{U^iU^j \over U^2}(\gamma-1) ,
\end{eqnarray}
\end{mathletters}
and $\gamma=1/\sqrt{1-U^2}$.  We expand the Lorentz transform
(\ref{ulorentz}) in powers of $U=|U^i|$, assuming that $U$ is of the
same order as $v_A$ so that the slow-motion assumption underlying the
post-Newtonian expansion is still valid.

We also account for the fact that the time slices in the primed coordinates
intersect the bodies' world lines at different points than in the unprimed
coordinates.  Without loss of generality, we can take the time slice in the
unprimed coordinates to be $t=0$ and the primed time slice to be $t'=0$.  Then
it follows from the inverse Lorentz transformation that quantities evaluated at
body $A$ at $t=0$ are evaluated at
\begin{equation}
t' = t'_A = y^{0'}_A(t=0) = -U^k y_A^k(t=0) .
\end{equation}
When transforming the equations of motion to the primed coordinates, we must
evaluate the primed vectors at $t'=0$ rather than $t'=t'_A$.  Therefore we
expand the transformed equations of motion in powers of the small quantity
$t'_A$.

Specifically, the position $y_1^i$ of body 1 on the time slice $t=0$
is related to the position $y_1^{i'}$ of body 1 on the slice $t'=0$ by
\begin{eqnarray}
y_1^i&\simeq&
y_1^{i'}-{1\over 2}U^i U^j y_1^{i'}-v_1^{i'}U^j y_1^{j'},
\end{eqnarray}
where $v_1^{i'}=dy_1^{i'}/dt'$.  Here and below, (unprimed) vectors on
the left-hand sides of equations are evaluated at $t=0$ while (primed)
vectors on right-hand sides are evaluated at $t'=0$.  By transforming
the four-velocity and four-acceleration, we find that
\begin{eqnarray}
v_1^i &\simeq& U^i \left[1 - {1\over 2}\left(U v_1'\right)
\right] + v_1^{i'} \left[1-{1\over 2}U^2 - \left(U v_1'\right) \right]
\nonumber\\&&
+ t'_1 {dv_1^{i'}\over dt'} ,
\\
{dv_1^i\over dt} &\simeq& \left[1 - U^2 - 2(Uv'_1)\right]
{dv_1^{i'}\over dt'} + t'_1 {d^2 v_1^{i'}\over dt^{'2}}
-v_1^{i'}U^k {dv_1^{k'}\over dt'}
\nonumber\\&&
-{1\over 2}U^i U^k {dv_1^{k'}\over dt'},
\label{e:accelorentz}
\end{eqnarray}
where $r_{12}' = |\bbox{y}'_1 - \bbox{y}'_2|$, $n_{12}^{i'} =
(y^{i'}_1 - y^{i'}_2)/r_{12}'$, and $t'_1 = \Lambda^0_\nu(-U)
y_1^\nu$.  In the same way, by considering the Lorentz transform and
time slicing of the spin variables and the spin supplementary
condition, we find that
\begin{eqnarray}
S_1^i&\simeq&
S_1^{i'}+{1\over 2}U^i (U S_1')+U^{i'}(S'_1 v'_1),
\\
S_1^{ij}&\simeq&S_1^{i'j'}\left[1-{1\over 2}U^2-(v'_1 U)\right]
+\varepsilon_{ijk}v_1^{k'}(S'_1 U) \nonumber\\
&&
+\varepsilon_{ijk}U^k[(v'_1 S'_1)+(U S'_1)]
-U^{[i'}S_1^{j']k'}U^k
\nonumber\\&&
-2U^{[i}S_1^{j']k'}v_1^{k'}.
\end{eqnarray}
The transformations of the variables of body 2 are given by exchanging
subscripts 1 and 2 in the above formulas.

We insert these formulas into the equations of motion
(\ref{e:eomall})--(\ref{e:eom2.5pnSO}), keeping the lowest-order
acceleration on the right hand side of Eq.~(\ref{e:accelorentz}) but
eliminating other (post-Newtonian) accelerations with the 1.5PN
equations of motion in the primed coordinates. Thus we obtain
equations of motion in the primed coordinates which, to 2.5PN order,
have the same form as Eq.~(\ref{e:eom2.5pnSO}). This completes the
check of the Lorentz invariance of our 2.5PN equations of motion.

\end{document}